\documentclass[10pt]{article}

\usepackage{latexsym}
\usepackage{graphics}
\usepackage{amsfonts}
\usepackage{undertilde}

\newcommand{\be}{\begin{equation}}
\newcommand{\ee}{\end{equation}}
\newcommand{\ben}{\begin{eqnarray}}
\newcommand{\een}{\end{eqnarray}}
\newcommand{\bea}{\begin{eqnarray}}
\newcommand{\eea}{\end{eqnarray}}

\begin{document}

 \setlength{\baselineskip}{17pt}
\title{
\normalsize
\mbox{ }\hspace{\fill}
\begin{minipage}{7cm}
{\tt }{\hfill}
\end{minipage}\\[5ex]
{\large\bf From Lagrangian to Hamiltonian formulations
 \\of the Palatini action
 \\[1ex]}}
\author{SangChul Yoon\footnote{scyoon@kunsan.ac.kr}
\\
\\
{\it Department of Physics, Kunsan National University},\\
{\it Kunsan 573-701, Korea}\\
}

\maketitle

\thispagestyle{empty}

\begin{abstract}
We work on the Lagrangian and the Hamiltonian formulations of the
Palatini action. In the Lagrangian formulation, we find that we need
to assume the metric compatibility and the torsion zero or to assume
the tetrad compatibility  to describe General Relativity. In the
Hamiltonian formulation, we obtain the Einstein's equations only
with assuming the tetrad compatibility. The Hamiltonian from
assuming the metric compatibility and the torsion zero should be
used to quantize General Relativity.
\end{abstract}

\section{Introduction}
The tetrad and the internal connection formulation of General
Relativity has been studied more than 30 years, yet it is still
obscure what should be assumed beforehand and what are derived
afterward from the Euler-Lagrange equations in the beginning
Lagrangian formulation of this program. In this paper, we clear this
up once and for all. This makes the Hamiltonian formulation more
interesting than previously known.

We derive the Palatini action from the Einstein-Hilbert action. From
the variational principle, we find that varying the connection, we
have the compatibility condition of the connection with the tetrad
when we assume the metric compatibility and the torsion zero
conditions. Varying the tetrad, we have the Einstein equations. When
the torsion is not zero, varying the connection gives us the torsion
zero condition if the connection is compatible with the tetrad. In
the Lagrangian formulation, we find these two approaches to describe
General Relativity, which we apply to the Hamiltonian formulation.

We perform the Legendre transformation and obtain the Hamiltonian.
There are 2nd class constraints. From the lesson above, we solve
these and obtain the scalar, vector and Gauss constraints. In the
first approach of the metric compatibility and the torsion zero
conditions, the Hamiltonian equations of motion are different from
the Einstein's equations. In the second approach of the tetrad
compatibility condition, the Hamiltonian equations of motion become
the Einstein's equations after solving the Gauss constraint.

In section 2, we introduce Riemannian geometry \cite{Lee}. Spacetime
and spatial tensor indices are denoted by the alphabet $a, b,
\cdot\cdot\cdot ,$ while internal indices are denoted by the
alphabet $i, j,  \cdot\cdot\cdot $ for 3-dimension and $I, J,
\cdot\cdot\cdot $ for 4-dimension. The signature of the spacetime
metric $g_{ab}$ is taken to be $(-+++)$.

\section{Connection and Torsion} Consider a 4-dimensional
manifold $M$, and let $V$ be a 4-dimensional vector space with
Minkowski metric $\eta_{IJ}$ having signature $(-+++)$. A tetrad at
$p\in M$ is an isomorphism $e^a_I(p):V\rightarrow T_pM$ and can act
on tensors. For example \bea \eta_{IJ}=g_{ab}e^a_I e^b_J. \eea The
inverse of $e^a_I$ will be denoted by $e_a^I$. It satisfies \bea
\eta_{IJ} e^I_a e^J_b=g_{ab}. \eea Spacetime tensor fields with
additional internal indices $I, J, \cdot\cdot\cdot$ will be called
generalized tensor fields on $M$. Spacetime indices are raised and
lowered with the spacetime metric $g_{ab}$; internal indices are
raised and lowered with the Minkowski metric $\eta_{IJ}$.

A generalized derivative operator obey the linearity, Leibnitz rule,
and commutativity with contraction  with respect to both the
spacetime and the internal indices. We require that all generalized
derivative operators be compatible with  $\eta_{IJ}$.   If
$\partial_a$ is a derivative operator, then any other generalized
derivative operator $D_a$ is defined  by a pair of generalized
tensor fields $A_{ab}^{\mbox{ }\mbox{ }c}$ and $w_{aI}^{\mbox{
}\mbox{ } J}$: \bea D_a H_{bI}\equiv
\partial_a H_{bI}+A_{ab}^{\mbox{ }\mbox{ }c} H_{cI} + w_{aI}^{\mbox{
}\mbox{ } J} H_{bJ}.\eea From $D_a \eta_{IJ}=0$, we obtain \bea
w_{aIJ}=w_{a[IJ]}. \eea If $D_a g_{bc}=0$, \bea A_{ab}^{\mbox{
}\mbox{ }c} = \Gamma_{ab}^{ \mbox{ } \mbox{ } c} + \frac{1}{2} \{
-T^{\mbox{ }c}_{a \mbox{ } \mbox{ }b}- T^{\mbox{ }c}_{b \mbox{
}\mbox{ }a} +T_{ab}^{ \mbox{ } \mbox{ } c} \}, \eea where $
\Gamma_{ab}^{ \mbox{ } \mbox{ } c}$ is the Christoffel symbols, \bea
\Gamma_{ab}^{ \mbox{ } \mbox{ } c} = -\frac{1}{2} g^{cd}\{
\partial_b g_{ad}+
\partial_a g_{bd} -\partial_d g_{ab} \} \eea and $T_{ab}^{ \mbox{ }
\mbox{ } c}$ is the torsion, \bea T_{ab}^{ \mbox{ } \mbox{ }
c}\equiv A_{ab}^{ \mbox{ } \mbox{ } c}-A_{ba}^{ \mbox{ } \mbox{ }
c}, \eea which measures the failure of the closure of the
parallelogram made up of small displacement vectors and their
parallel transports \cite{Nakahara} and the non-commutativity of the derivative
operator on a scalar field $f$ such that \bea
D_aD_bf-D_bD_af=T_{ab}^{ \mbox{ } \mbox{ } c}D_c f.\eea  If
$T_{ab}^{ \mbox{ } \mbox{ } c}=0$, just as a compatibility with a
spacetime metric $g_{ab}$ defines a unique, torsion-free spacetime
derivative operator, compatibility with $e^a_I$  defines a unique
torsion-free generalized derivative operator $\nabla _a$ defined by
\bea \nabla _a e_{bI}\equiv
\partial_a e_{bI}+\Gamma_{ab}^{\mbox{ }\mbox{ }c} e_{cI}+
w_{aI}^{\mbox{ }\mbox{ } K} e _{bK}=0.\eea Whether  the torsion is
zero or not, the compatibility condition gives  \bea w_{aI}^{\mbox{
}\mbox{ } J} = - e^{bJ}(\partial_a e_{bI}+A_{ab}^{\mbox{ }\mbox{ }c}
e_{cI}). \eea In this case, $w_{aI}^{\mbox{ }\mbox{ } J}$ is the
spin connection. It  is related to the spacetime geometry and has
informations about the torsion and the curvature.

In the notation of differential form, the torsion is defined as \bea
\bold{T}^I\equiv d\bold{e}^I+\bold{w}^I_{\mbox{ } J} \wedge
\bold{e}^J \quad{ } \quad{ } \eea which means \bea T_{ab}^{\mbox{
}\mbox{ }I}= 2\partial_{[a}e_{b]}^I +w_{a \mbox{ } J}^{\mbox{ } I}
e_b^J-w_{b \mbox{ } J}^{\mbox{ } I} e_a^J.\eea In Riemannian
geometry, (10) is always satisfied. In this case $T_{ab}^{\mbox{
}\mbox{ }I}$ and $T_{ab}^{\mbox{ }\mbox{ }c}$ are equivalent: \bea
T_{ab}^{\mbox{ }\mbox{ }c}=T_{ab}^{\mbox{ }\mbox{ }I}e^c_I.\eea For
the zero torsion, we can write $w_{aI}^{\mbox{ }\mbox{ } J}$ in
terms of $e^a_{I}$ using (12) and it turns out to be equivalent to
(10) with $A_{ab}^{\mbox{ }\mbox{ }c}=\Gamma_{ab}^{\mbox{ }\mbox{
}c}$. For the non-zero torsion, if we plug (10) into (12), we obtain
(7). In the connection formulation of the Palatini action, $e^a_I$
and $w_{aI}^{\mbox{ }\mbox{ } J}$  are the basic independent
variables. Therefore (10) and (13) are not satisfied in general and
(12) does not have the geometrical meanings of Riemannian geometry.

Given a generalized derivative operator $D_a$, we can construct
curvature tensors by commuting derivatives. For the torsion zero,
the internal curvature tensor $F_{abI}^{ \mbox{ } \mbox{ } \mbox{ }
J}$and the spacetime curvature tensor $\tilde{F}_{abc}^{ \mbox{ }
\mbox{ } \mbox{ } d}$ are defined by \bea 2D_{[a}D_{b]}H_I \equiv
F_{abI}^{ \mbox{ } \mbox{ } \mbox{ } J}H_J \quad \mbox{and}\eea \bea
2D_{[a}D_{b]}H_c \equiv \tilde{F}_{abc}^{ \mbox{ } \mbox{ } \mbox{ }
d}H_d. \eea From these \bea F_{abI}^{ \mbox{ } \mbox{ } \mbox{ }
J}=2\partial_{[a}w_{b]I}^{\mbox{ } \mbox{ }J} +[w_a, w_b]_I^{\mbox{
} J} \quad \mbox{and } \eea \bea \tilde{F}_{abc}^{ \mbox{ } \mbox{ }
\mbox{ } d}=2\partial_{[a}A_{b]c}^{\mbox{ } \mbox{ }d} +[A_a,
A_b]_c^{\mbox{ } d}. \eea Here $[w_a, w_b]_I^{\mbox{ } J}=
(w_{aI}^{\mbox{ }\mbox{ } K}w_{bK}^{\mbox{ }\mbox{ }
J}-w_{bI}^{\mbox{ }\mbox{ } K}w_{aK}^{\mbox{ }\mbox{ } J})$ and
$[A_a, A_b]_c^{\mbox{ } d} = (A_{ac}^{\mbox{ }\mbox{
}e}A_{be}^{\mbox{ }\mbox{ }d}-A_{bc}^{\mbox{ }\mbox{
}e}A_{ae}^{\mbox{ }\mbox{ }d})$. For the non-zero torsion, we have
an additional term from the torsion to keep the linearity of the
curvature tensor \cite{Penrose and Rindler} \bea (2D_{[a}D_{b]}-T_{ab}^{\mbox{ }\mbox{ }c}D_c)
H_I \equiv F_{abI}^{ \mbox{ } \mbox{ } \mbox{ } J}H_J \quad
\mbox{and}\eea \bea (2D_{[a}D_{b]}-T_{ab}^{\mbox{ }\mbox{ }d}D_d)
H_c \equiv \tilde{F}_{abc}^{ \mbox{ } \mbox{ } \mbox{ } d}H_d .\eea

 We denote internal  and spacetime curvature tensors of the unique
 torsion-free generalized derivative operator $\nabla_a$ by $R_{abI}^{ \mbox{ } \mbox{ } \mbox{ }
 J}$ and $R_{abc}^{ \mbox{ } \mbox{ } \mbox{ } d}$. From (14) and (15), we can see that they are
 related by
 \bea R_{abI}^{ \mbox{ } \mbox{ } \mbox{ }
 J}= R_{abc}^{ \mbox{ } \mbox{ } \mbox{ } d}e^c_I e^J_d.\eea

\section{Palatini theory: Lagrangian formulation}

The Einstein-Hilbert action is \bea S_{EH}(g^{ab})= \int_M
\sqrt{-g}R \eea and \bea \sqrt{-g}R&=&
\sqrt{-g}\delta^c_{[e}\delta^d_{f]}R_{cd}^{ \mbox{ } \mbox{ } ef}
\nonumber \\
&=&-\frac{1}{4}\tilde{\eta}^{abcd}\epsilon_{abef}R_{cd}^{ \mbox{ }
\mbox{ } ef} \nonumber \\
&=& -\frac{1}{4}\tilde{\eta}^{abcd} \epsilon_{IJKL}e^I_a e^J_b e^K_e
e^L_f R_{cd}^{ \mbox{ } \mbox{ } ef} \nonumber \\
&=&-\frac{1}{4}\tilde{\eta}^{abcd} \epsilon_{IJKL}e^I_a e^J_b
R_{cd}^{ \mbox{ } \mbox{ } KL} \eea where $\tilde{\eta}^{abcd}$ is
the Levi-Civita tensor density of weight 1 and \bea
\epsilon_{abcd}=\epsilon_{IJKL}e^I_a e^J_b e^K_c e^L_d, \eea which
relates the volume element $\epsilon_{abcd}$ of $g_{ab}$ to the
volume element $\epsilon_{IJKL}$ of $\eta_{IJ}.$ The
Einstein-Hilbert action in terms of a co-tetrad $e^I_a$ is  \bea
S_{EH}(e)= -\frac{1}{4} \int_M \tilde{\eta}^{abcd}
\epsilon_{IJKL}e^I_a e^J_b  R_{cd}^{ \mbox{ } \mbox{ } KL}. \eea

In the Palatini action, $e^a_I$ and $w_{a}^{ \mbox{ } IJ}$ are the
basic independent variables. By replacing $R_{abI}^{ \mbox{ } \mbox{
} \mbox{ } J}$ in (24) with the internal curvature tensor $F_{abI}^{
\mbox{ } \mbox{ } \mbox{ } J}$ of an arbitrary generalized
derivative operator $D_a$ defined by (3), we obtain the 3+1 Palatini
action based on $SO(3,1)$: \bea S_p(e,w) \equiv -\frac{1}{8}\int_M
\tilde{\eta}^{abcd} \epsilon_{IJKL}e^I_a e^J_b F_{cd}^{ \mbox{ }
\mbox{ } KL}. \eea   An additional factor 1/2 which will not affect
the Euler-Lagrange equations of motion is included for the
Hamiltonian formulation. With \bea \tilde{\eta}^{abcd}
\epsilon_{IJKL}e^K_c e^L_d=-4\sqrt{-g}e^{[a}_I e^{b]}_J,\eea the
Palatini action is \bea S_p(e,w) \equiv \frac{1}{2}\int_M \sqrt{-g}
e^a_I e^b_J F_{ab}^{ \mbox{ } \mbox{ } IJ}. \eea $\sqrt{-g}$ is the
determinant of a metric $g_{ab}$, which is the determinant of
$e_a^I$ from $g_{ab}=\eta_{IJ}e_a^I e_b^J$. Because we are
interested in the role of the metric compatibility condition, this
expression $\sqrt{-g}$ here is useful.

It is also important to write the exact statement of a relation
between the metric compatibility condition, the torsion zero
condition and the tetrad compatibility condition: If $D_ag_{bc}=0$
and $T_{ab}^{\mbox{ }\mbox{ }c}=0$, then
 $T_{ab}^{\mbox{ }\mbox{ }I}=0$ if and only if $D_a e^b_I=0$ \cite{Peldan}.
Stokes's theorem holds for
 a torsion-free derivative operator on a orientable manifold and
 Gauss's theorem holds  when the metric compatibility condition is satisfied once
a volume element is chosen by a metric. Because great care must
 be taken to apply the variational principle without $D_ag_{bc}=0$ or the torsion
 zero condition, let's work on a simple model first:
 \bea S \equiv \int_M \sqrt{-g} P^a Q^b
D_a R_b. \eea If $D_ag_{bc}=0$ and $T_{ab}^{\mbox{ }\mbox{ }c}=0$,
\bea
\partial_a(\sqrt{-g}P^a)=\sqrt{-g} D_a P^a \eea  where we used the
formula: \bea \partial_a \sqrt{-g}=\frac{1}{2}\sqrt{-g}g^{bc}
\partial_a g_{bc} =-\sqrt{-g}(A_{ba}^{\mbox{ }\mbox{ }b}-T_{ba}^{\mbox{ }\mbox{
}b}) .\eea Note that the first equality holds also for $D_a$ and we
have \bea D_a\sqrt{-g}=\partial_a\sqrt{-g}+A_{ab}^{\mbox{ }\mbox{
}b}\sqrt{-g}. \eea  Generally without assuming $D_ag_{bc}=0$, \bea
\partial_a(\sqrt{-g}P^a) = \sqrt{-g} D_a P^a+( \partial_a\sqrt{-g} +\sqrt{-g}
 A_{ba}^{\mbox{ }\mbox{ }b})P^a=D_a(\sqrt{-g} P^a)+\sqrt{-g}T_{ba}^{\mbox{
}\mbox{ }b}P^a. \eea

   Let's see what we have when we
vary $R_a$. From $\delta S=0$, we have   \bea \sqrt{-g} D_a (P^aQ^b)
+( \partial_a\sqrt{-g} +\sqrt{-g} A_{ca}^{\mbox{ }\mbox{ }c})P^aQ^b
=0,\eea where we used $\delta R_a=0$ on the boundary. Note that the
second term does not disappear as in (32). If $T_{ab}^{\mbox{
}\mbox{ }c}=0$, we have \bea D_a (\sqrt{-g}P^aQ^b)=0. \eea We can
see that integration by parts works for $D_a$ when $T_{ab}^{\mbox{
}\mbox{ }c}=0$. A solution $D_a (P^aQ^b)=0$ is obtained only when
$D_ag_{bc}=0$ and $T_{ab}^{\mbox{ }\mbox{ }c}=0$.

Let's work on the Palatini action with the variational method. To
see what we have when we vary $e_I^a$, note that
$\tilde{\eta}^{abcd}$ and $\epsilon_{IJKL}$  are -1 or 0 or 1
depending on their indices, so they are independent of $e_I^a$. With
this, varying $e_I^a$ in (25) gives \bea \tilde{\eta}^{abcd}
\epsilon_{IJKL} e^J_b F_{cd}^{ \mbox{ } \mbox{ } KL}=0. \eea For
$w_{a}^{ \mbox{ } IJ}$, we need the following formula: \bea \delta
F_{ab}^{ \mbox{ } \mbox{ } IJ}=2D_{[a}\delta w_{b]}^{\mbox{ }
IJ}-T_{ab}^{\mbox{ }\mbox{ }c} \delta w_{c}^{\mbox{ }IJ}. \eea We
can see immediately that the variational calculations of the
Palatini action (27) with respect to $w_{a}^{ \mbox{ } IJ}$ are very
similar to those of our simple action (28).

If we assume $D_ag_{bc}=0$ and $T_{ab}^{\mbox{ }\mbox{ }c}=0$,
varying $w_{a}^{ \mbox{ } IJ}$ gives us  \bea D_a( e^{[a}_I
e^{b]}_J)=0. \eea To determine what (37) gives, let us express $D_a$
in terms of the unique, torsion-free generalized derivative operator
$\nabla_a$ compatible with $e^I_a$, and $C_{aI}^{\mbox{ }\mbox{ }
J}$ \cite{Romano} defined by \bea
D_aH^b_I=\nabla_aH^b_I+C_{aI}^{\mbox{ }\mbox{ } J}H^b_J.\eea Note
that this expression is possible only when $D_ag_{bc}=0$.
Multiplying $e^J_b$ to (37), we have $ D_a e^a_I= 0$. Since $e^a_I$
is invertible, combining these we get \bea e^a_I
C_{aJK}-e^a_JC_{aIK}=0. \eea Multiplying $e^I_b$, \bea C_{bJK}=e^a_J
e^I_b C_{aIK}. \eea With $C_{bJK}+C_{bKJ}=0$, we have \bea e^a_J
C_{aIK}+e^a_KC_{aIJ}=0. \eea With index substitutions $I \rightarrow
K, J \rightarrow I, K \rightarrow J$, \bea e^a_I
C_{aJK}+e^a_JC_{aIK}=0. \eea With (39) and (42), we obtain
$C_{aIJ}=0$.  Algebraically there are 24 homogeneous linear
equations of 24 variables $C_{aI}^{\mbox{ }\mbox{ } J}$, so
$C_{aI}^{\mbox{ }\mbox{ } J}=0$. Since $C_{aI}^{\mbox{ }\mbox{ }
J}=0$, we find that one equation of motion implies that
$D_a=\nabla_a$ and $F_{ab}^{ \mbox{ } \mbox{ } IJ}=R_{ab}^{ \mbox{ }
\mbox{ }I J}$. The remaining Euler-Lagrange equation of motion
becomes \bea \tilde{\eta}^{abcd} \epsilon_{IJKL} e^J_b  R_{cd}^{
\mbox{ } \mbox{ } KL}=0. \eea When (43) is contracted with $e^{eI}$,
we get the 3+1 vacuum Einstein's equation, $G^{ae}=0$.

If we do not assume $D_ag_{bc} = 0$ but only assume $T_{ab}^{\mbox{
}\mbox{ }c}=0$, we have  \bea \sqrt{-g} D_a ( e^{[a}_I e^{b]}_J)
+(\partial_a\sqrt{-g} +\sqrt{-g} A_{ca}^{\mbox{ }\mbox{ }c})(
e^{[a}_I e^{b]}_J) =0.\eea In this case,  we need to add
$B_{ab}^{\mbox{ }\mbox{ } c}=A_{ab}^{\mbox{ }\mbox{ }
c}-\Gamma_{ab}^{\mbox{ }\mbox{ } c}$ in (38) to determine what (44)
gives such that \bea D_aH^b_I=\nabla_aH^b_I-B_{ac}^{\mbox{ }\mbox{ }
b}H^c_I +C_{aI}^{\mbox{ }\mbox{ } J}H^b_J.\eea If we express (44)
with (45), there are 24 inhomogeneous linear equations of 24
variables $C_{aI}^{\mbox{ }\mbox{ } J}$, so $C_{aI}^{\mbox{ }\mbox{
} J} \neq 0$. In this case,  $D_a e^b_I$ is not zero. The Palatini
action does not become the Einstein-Hilbert action and we do not
have the Einstein's equations. If we assume 40 components of $D_a
e^b_I$ are zero, which are linear relations between $B_{ab}^{\mbox{
}\mbox{ } c}$ and $C_{aI}^{\mbox{ }\mbox{ } J}$, we obtain other 24
components of $D_a e^b_I$ are zero from (44). However, this
assumption is not  covariant.  Therefore we must assume $D_ag_{bc} =
0$.

Finally if we do not assume $T_{ab}^{\mbox{ }\mbox{ }c}=0$, varying
$w_{a}^{ \mbox{ } IJ}$ gives us \bea 2D_a( \sqrt{-g}e^{[a}_I
e^{b]}_J) +\sqrt{-g}(2e^a_{[I}e^b_{J]}T_{ca}^{\mbox{ }\mbox{ }c}+
e^a_Ie^c_JT_{ac}^{\mbox{ }\mbox{ }b})=0.\eea If we assume
$D_ae^b_I=0$, we multiply $e^J_b$ to both sides and obtain
$T_{ac}^{\mbox{ }\mbox{ }c}=0$. Thus we have $T_{ab}^{\mbox{ }\mbox{
}c}=0$ and the Palatini action describe General Relativity.

Since $D_ae^b_I=0$ means $D_ag_{bc}=0$,  we can see that we must
assume $D_ag_{bc}=0$ to have the Einstein's equations from the
Palatini action. Because this condition is assumed from the
beginning, it must be preserved in quantization. We also need to
assume either $T_{ab}^{\mbox{ }\mbox{ }c}=0$ or $D_ae^b_I=0$ to have
the Einstein's equations, which should be also preserved in
quantization. The conditions  $D_ag_{bc}=0$ and $T_{ab}^{\mbox{
}\mbox{ }c}=0$ are what Einstein assumed  when he constructed
General Relativity \cite{Ohanian and Ruffini}. With these two
conditions, geodesic is a extremal length between two spacetime
points, which is related to the Principle of Equivalence. On the
other hand, assuming $D_ae^b_I=0$ is based on Riemannian geometry.
It is straightforward to check that our results also hold for the
Holst action \cite{Holst}.

\section{Palatini theory: Hamiltonian formulation}

Before working on the Hamiltonian formulation of the Palatini
action, let's discuss the equivalence of the Lagrangian and the
Hamiltonian formulation. To construct the Hamiltonian, we define the
momentum variable $p_i$ from the Lagrangian $L(q_i, \dot{q}_i)$:
\bea p_i=\frac{ \partial L(q, \dot{q})}{ \partial \dot{q}_i}. \eea
We obtain the Hamiltonian with the Legendre transformation: \bea
H(q,p)=\dot{q}_i p_i - L(q,\dot{q}). \eea With this,  we obtain the
Hamiltonian equations of motion: \bea \dot{q}_i=\frac{\partial
H}{\partial p_i}, \eea \bea \dot{p}_i=-\frac{\partial H}{\partial
q_i}. \eea The Euler-Lagrange equations are equivalent to the
Hamiltonian equations when (47) is equivalent to (49). In the
Palatini action, the independent variables are $e^a_I$ and $w_{a}^{
\mbox{ } IJ}$. $\dot{q}_i$ of $w_{a}^{ \mbox{ } IJ}$ comes from
$F_{abI}^{ \mbox{ } \mbox{ } \mbox{ } J}$, but where is $\dot{q}_i$
of $e^a_I$? Because the metric compatibility condition and the
torsion zero condition deal only with $A_{ab}^{\mbox{ }\mbox{ }c}$,
$\dot{q}_i$ of $e^a_I$ comes from the tetrad compatibility
condition. Therefore we will see that only in the second approach,
the Lagrangian and the Hamiltonian formulation are equivalent. The
Hamiltonian equations of motion from the first approach should be
treated as one of modifications of General Relativity for
quantization \cite{Dirac-Feynman}.

Let's work on the Hamiltonian formulation of the first approach. To
perform the Legendre transformation, we introduce a foliation
 $\{\Sigma\}$ in space-time and a time-like vector field $t^a$ whose
integral curves intersect each $\Sigma$ of the foliation precisely
once. Let $n^a$ denote the unit normal to the foliation. We can then
decompose the time-evolution vector field $t^a$ normal and
tangential to the foliation: \bea t^a=Nn^a+N^a, \quad n^aN_a=0. \eea
The function $N$ is called the lapse function and the vector field
$N^a$ is called the shift vector \cite{Wald}. Given $n^a$, it
follows that $q^a_b=\delta^a_b+n^a n_b$ is a projection operator
into the foliation. Let $E^a_I=q^a_b e^b_I$. Let $ ^4D $ denote an
derivative operator on $M$ and $D_a=q_a^b {^4}D_b$ on $\Sigma $. We
can now decompose the action (27): \bea e^a_I e^b_J F_{ab}^{\mbox{
}\mbox{ } IJ}= e^a_I e^b_J (q^c_a-n^c n_a)(q^d_b- n^d n_b)
F_{cd}^{\mbox{ }\mbox{ } IJ}. \eea The first term becomes
$E^a_IE^b_J F_{ab}^{\mbox{ }\mbox{ } IJ}$, the last term becomes
zero and the cross terms are: \bea -2e^a_Ie^b_J q^c_a n^d n_b
F_{cd}^{\mbox{ }\mbox{ } IJ}&=& -2E^c_I n_J (
\frac{1}{N}t^d-\frac{N^d}{N}) F_{cd}^{\mbox{ }\mbox{ } IJ} \nonumber
\\ &=& -2E^c_In_J \frac{ -\mathcal{L}_t w_c^{\mbox{ } IJ} + {^4D}_c(t^d
w_d^{\mbox{ } IJ})}{N}+ 2\frac{N^d}{N}E^c_In_J F_{cd}^{\mbox{
}\mbox{
} IJ}, \nonumber \\
   \eea
where $t^d F_{cd}^{\mbox{ }\mbox{ } IJ}=-\mathcal{L}_t w_c^{\mbox{ }
IJ}+{^4}D_c(t^d w_d^{\mbox{ } IJ})$. The action becomes: \bea S_p=
\frac{1}{2} \int dt \int d^3x \sqrt{q} \Big( N E^a_I E^b_J
F_{ab}^{\mbox{ }\mbox{ } IJ} +2 n_{[I}E^a_{J]} D_a(w \cdot
t)^{\mbox{ } IJ}\nonumber
\\-2n_{[I}E^a_{J]} \dot{w}_a^{\mbox{ }
IJ}+2N^an_{[I}E^b_{J]}F_{ab}^{\mbox{ }\mbox{ } IJ}\Big), \eea where
$\sqrt{-g}=N\sqrt{q}$. Note that all $ a,b \cdot\cdot\cdot$ are
spatial and now $F_{ab}^{\mbox{ }\mbox{ } IJ}$ in (54) is the
curvature tensor of $D_a$. To further simplify the action, we define
\bea  \tilde{E}^a_I\equiv \sqrt{q}E^a_I, \eea \bea
\tilde{E}^a_{IJ}\equiv \tilde{E}^a_{[I}n_{J]}.\eea Because there is
not much confusion, we keep using $ \tilde{E}$ for
$\tilde{E}^a_{IJ}$. With this \bea \mbox{tr}( \tilde{E}^a
\tilde{E}^b F_{ab}) &=& \tilde{E}^a_{[I}n_{J]} \tilde{E}^{b[J}n^{K]}
F_{abK}^{ \mbox{ } \mbox{ } \mbox{ } I}\nonumber\\ &=& \frac{1}{4}
\tilde{E}^a_I \tilde{E}^b_K F_{ab}^{ \mbox{ } \mbox{ } K I},
 \eea
where we have used that $E^a_I n^I=0$. In this way, the action
becomes: \bea S_p= \int dt \int d^3 x \mbox{tr} \Big( - N \frac{2}{\sqrt{q}}
\tilde{E}^a \tilde{E}^b F_{ab}+N^a\tilde{E}^bF_{ab}-(w \cdot t)D_a\tilde{E}^a-\tilde{E}^a\dot{w}_a\Big), \eea where we used the
the fact that the torsion zero condition in 4-dimension makes the torsion in 3-dimension vanish.

We can see that $w_a^{\mbox{ } IJ}, -\tilde{E}^a_{IJ}$ are canonical
variables and $N,N^a, (w \cdot t)^{\mbox{ } IJ}$ are
non-dynamical. They serve as Lagrange multipliers. Variation of the
action with respect to these fields yields the constraints: \bea H_s
\equiv \frac{2}{\sqrt{q}}\mbox{tr}(\tilde{E}^a \tilde{E}^b F_{ab} )\approx 0,\eea
\bea V_a\equiv -\mbox{tr}(\tilde{E}^bF_{ab})\approx 0, \eea \bea
G_{IJ} \equiv -D_a \tilde{E}_{IJ}^a \approx 0. \eea    The
Hamiltonian up to surface terms is \bea H=\int_\Sigma d^3 x \Big(
NH_s+N^aV_a+(t \cdot w)^{\mbox{ } IJ} G_{IJ}\Big).\eea

There are second class constraints in this formulation. Not all $\tilde{E}^a_{IJ}$ are
independent and we have a primary constraint \bea \phi^{ab}\equiv
\epsilon^{IJKL} \tilde{E}^a_{IJ}\tilde{E}^b_{KL} \approx 0,  \eea
which is obvious from (56). All Poisson brackets between constraints
vanish weakly except one between $H_s$ and $\phi^{ab}$
\cite{Ashtekar} . The secondary constraints from this is \bea
\chi^{ab} \equiv \epsilon^{IJKL}D_c\tilde{E}^a_{IJ}
[\tilde{E}^b,\tilde{E}^c]_{KL}+(a\leftrightarrow b) \approx 0, \eea
where
$[\tilde{E}^b,\tilde{E}^c]_{KL}=\tilde{E}^b_{KN}\tilde{E}^{cN}_{\mbox{
} \mbox{ } L}-\tilde{E}^c_{KN}\tilde{E}^{bN}_{\mbox{ } \mbox{ } L}.
$ The Poisson bracket between $\chi^{ab}$ and the total Hamiltonian
vanishes weakly, and \bea \{\phi^{ab}(x), \chi^{cd}(y)\}\approx
8q(2q^{ab}q^{cd}-q^{ad}q^{bc}-q^{ac}q^{bd})\neq0.\eea Thus we do not
have any more constraints and $\phi^{ab},\chi^{ab}$ are the second
class constraints.

Now how to solve the second class constraints? We have learned from
the Lagrangian formulation of the Palatini theory that we need the
tetrad compatibility condition to have the Einstein Equation. For
the Hamiltonian formulation, we break 4-dimensional diffeomorphic
covariance to 1+3, but we still have 3-dimensional covariance.
Therefore we might guess that the 3-dimensional triad compatibility
condition  can solve the 2nd class constraints. We will see that
this turns out to be the case.

To solve (64), we fix $n_I$ by $\partial_a n_I =0 $.
This makes an internal vector field $n_I$ become an internal vector, which means we break 4-dimensional
internal covariance to 3+1. With this,
$\tilde{E}^a_{IJ}$ has 9 degrees of freedom from $\tilde{E}^a_I$.
To make $w_a^{\mbox{ } IJ}$  also have 9 degrees of freedom, we also request
\bea n^I G_{IJ}=0, \eea because (64) has only 6 components, which are equations of
$w_a^{\mbox{ } IJ}$  with only spatial $I$ and $J$.  To solve (64) and (66),
we express $D_a$ in terms of the unique, torsion-free generalized derivative operator
$\nabla_a$ compatible with $E^a_I$, and $c_{aI}^{\mbox{ }\mbox{ }
J}$ defined by \bea D_aH^b_I=\nabla_aH^b_I+c_{aI}^{\mbox{ }\mbox{ }
J}H^b_J.\eea This is possible from the metric compatibility assumption.
(64) and (66) are 9 independent homogeneous equations of $c_{aI}^{\mbox{ }\mbox{ }
J}$ with  spatial $I$ and $J$, so it is zero. Thus $w_{aI}^{\mbox{ }\mbox{ }
J}$ with spatial $I,J$ is the spin connection which is completely determined by $\tilde{E}^a_I$.
   Because the boost part of $w_{aI}^{\mbox{ }\mbox{ } J}$ is free, we can
write as \bea w_a^{\mbox{ } IJ}=\Gamma_a^{\mbox{ } IJ}+
2K_a^{[I}n^{J]}, \eea where $\Gamma_a^{\mbox{ } IJ}$ is the spin
connection on $\Sigma$ and $K_a^In_I=0$.  Because $\tilde{E}^a_I
n^I=0$ also, we will use 3-dimensional internal index $i$ and write
these variables as $(K^i_a,\tilde{E}^a_i)$. Thus, after eliminating
the 2nd class constraints and fixing $n_I$, the phase space of the
Palatini theory is  the pair $(K^i_a,\tilde{E}^a_i)$ and the only
non-vanishing Poisson bracket is \bea \{
K^i_a(x),\tilde{E}^b_j(y)\}=\delta^b_a \delta^i_j \delta^3(x,y).
\eea Starting from 16 components of $e^a_I$, 40 of $A_{ab}^{ \mbox{
} \mbox{ } c}$ and 24 $w_{a}^{ \mbox{ } IJ}$, we are left with  18
degrees of freedom  by 40 of $D_a g_{bc}=0$, 3 of $\partial_a
n_I=0$, 9 of $\Gamma_{a}^{ \mbox{ } IJ}$ and 10 non-dynamical $N,
N^a, (w \cdot t)^{ IJ}$. With the 7 first class constraints, we have
2 degrees of freedom \cite{Wald}.

Finally let's write down the 7 first class constraints with this
pair. It is straightforward if we write down $F_{ab}^{\mbox{ }\mbox{
} IJ}$ using (68): \bea
H_s=-\frac{1}{2}\sqrt{q}\mathcal{R}-\frac{1}{\sqrt{q}}\tilde{E}^a_{[i}\tilde{E}^b_{j]}
K^i_a K_b^j \approx 0, \eea \bea V_a=-2\tilde{E}^b_i D_{[a}K^i_{b]}
\approx 0,\eea \bea G_{ij}=\tilde{E}^a_{[i}K_{aj]} \approx 0, \eea
where $\mathcal{R}$ denotes the scalar curvature of $D_a$ which is
the unique torsion-free derivative operator compatible with $E^a_i$.
We will call (70), (71), and (72) the scalar, vector, and Gauss
constraints. If ${^4}D_ae^b_I=0$, $-K_{ab}$ is an extrinsic
curvature:
\bea E^i_bK_{ai}=e^I_b K_{aI}&=&-e^I_b q^c_a{^4}D_cn_I \nonumber \\
&=& -q^c_a {^4}D_cn_b \quad \mbox{if} {^4}D_ae^I_b=0. \eea Because
$K_{ab}=K_{ba}$, $G_{ij}=0$ is automatically satisfied. In this
case, (70) and (71) become the the scalar and  vector constraints of
the standard Einstein-Hilbert action. However $K^i_a$ is not the extrinsic curvature
because we do not assume ${^4}D_ae^b_I=0$. We will see that the Hamiltonian
formulation of the 3+1 Palatini theory in this approach is not the metric
description of General Relativity.

Suppose we start with the metric compatibility, the torsion zero and
the 3-dimensional triad compatibility conditions with fixing $n_I$. Then
there is no 2nd class constraint. This method can be applied to the Holst action
 and we obtain the phase space variables and the constraints of
Loop Quantum Gravity, which are originally derived by the canonical
transformation from $(K^i_a,\tilde{E}^a_i)$ \cite{Barbero}.

So far we have solved the second class constraints assuming the
metric compatibility and the torsion zero with fixing $n^I$. The
other approach is to assume the tetrad compatibility condition.
Here more second class constraints come from (36),
which are solved by the torsion zero on $\Sigma$. We can solve (64)
with a more covariant way directly from our assumption ${^4}D_ae^b_I=0$  with some care
because \bea D_a E^b_I &=& q_a^c q_d^{b4}D_cE^d_I \nonumber \\
                       &=& q_a^c q_d^b e_I^{e4}D_cq^d_e  \quad
                       \mbox{with}^4 D_a e^b_I=0 \nonumber \\
                       &=& q_a^c q_d^{b4} D_cn^d  n_I \quad
                       \mbox{with}^4D_ag_{bc}=0 \nonumber \\
                       &=&\bar{K}_a^bn_I,          \eea
which is not zero, and $\bar{K}_{ab}$ is the extrinsic curvature.
Therefore we need to use \bea
 q^I_J D_a E^b_I=0, \eea
where $q^I_J=q^a_bE_a^IE^b_J$.  It is straightforward to check that
(75) solves (64): \bea
\chi^{ab}=4\epsilon^{IJKL}\tilde{E}^a_I\tilde{E}^b_K\tilde{E}^c_LD_cn_J+(a\leftrightarrow
b)=0. \eea Furthermore only $q^I_K q^J_L G_{IJ}$ survives: \bea
            n^Iq^J_K G_{IJ}=0. \eea
As we mentioned, $\phi^{ab} = 0$ automatically by our construction.

Now we have 7 first class constraints.  In the same way as the first
approach, we fix $n_I$ by $\partial_a n_I =0 $. With this,
$\tilde{E}^a_{IJ}$ has 9 degrees of freedom from $\tilde{E}^a_I$.
$w_a^{\mbox{ } IJ}$ also has 9 degrees of freedom with spatial $I,J$
becoming the spin connection because (75) becomes the triad
compatibility condition on $\Sigma$. Therefore the phase space and
the constraints are the same with those of the first approach.
Because $K_{ab}$ is the extrinsic curvature from (73), this approach
is the metric description of General Relativity. To see what is
going on more clearly, we parameterize the foliation $\{\Sigma_t\}$
by a global time function $t$ which is possible if $M$ is  globally
hyperbolic. We also pick up a coordinate $\{x^{\mu}\}$ on $\Sigma$.
Let $t^a$ in (51) satisfy $t^a \nabla_a t =1$ and $t^a \nabla_a
x^{\mu}$=0. Let $N^a$ satisfy $N^a \nabla_a t$=0 and  $N^a \nabla_a
x^{\mu}=N^{\mu}$. In this coordinate, $t^a= (1,0)$,
$N^a=(0,N^{\mu})$ and $n^a=(1/N,-N^{\mu}/N)$. From (1),
$e^a_0=(e^t_0,e^{\mu}_0)$ and $e^a_i=(e^t_i,e^{\mu}_i)$ are
orthonormal vectors. If we choose $e^a_0=n^a$, then
$e^a_i=(0,E^a_i)$ by $\partial_a n_I=0$ \cite{Holst}.
 To make $q_a^{c}$$ ^4 D_cn_b$ symmetric with (a,b), we need
$q_a^c q_b^d T_{cd}^{\mbox{ }\mbox{ } e}=0$. We can easily see that this comes from the Gauss constraint.
Thus by solving the Gauss constraint, this approach becomes the metric description of General Relativity.

Finally let's come back to the first approach and write down the Hamiltonian equations of motion:
\bea \dot{\tilde{E}^a_i} &=&\{\tilde{E}^a_i,H\} \nonumber \\
                         &=& N(\tilde{E}^a_iK-\tilde{E}^b_iK_b^a)-\tilde{E}^a_iD_bN^b
+\tilde{E}^b_iD_bN^a+(t\cdot w)_i^j\tilde{E}^a_j, \\
 \dot{K}_a^i&=&\{K_a^i,H\} \nonumber \\
            &=& -N(R_a^i+KK_a^i-K_a^bK_b^i)+N^bD_aK_b^i-N^bD_bK_a^i+(t \cdot w)^{ij}K_{aj},\eea
where $R_{ab}$ is the Ricci tensor on $\Sigma$ and we impose the
triad compatibility condition after functional derivatives.

\section{Conclusion}
In the Lagrangian formulation of the Palatini action, we found that
there are two approaches to describe General Relativity. One is to
assume the metric compatibility and the torsion zero conditions and
the other is to assume the tetrad compatibility condition. In the
Hamiltonian formulation, we found that only the second approach
describes General Relativity. This is the metric description which
is very hard to quantize.

In the first approach of the metric compatibility and the torsion
zero assumptions, the time evolution of the tetrad is different from
that of General Relativity. This is a very unexpected result. We do
not know whether this has any meaning classical mechanically because
General Relativity is a established theory with experiments. We will
see what it means to quantized General Relativity with this
modification.

\end{document}